\documentclass[aps,prb,preprint]{revtex4-2}
\usepackage{amsmath,amssymb,mathrsfs,braket,bm,txfonts,titlesec}
\usepackage{graphicx}
\usepackage{hyperref}
\newcommand{\mykappa}{\eta}
\DeclareMathAlphabet{\mymathbb}{U}{BOONDOX-ds}{m}{n}
\begin{document}
\title{Massive Klein Tunneling in Topological Photonic Crystals}
\author{Keiji Nakatsugawa$^{1}$}
\author{Xiao Hu$^{1}$}
\email[Corresponding Author's Email: ]{HU.Xiao@nims.go.jp}
\affiliation{$^{1}$Research Center for Materials Nanoarchitectonics (MANA), National Institute for Materials Science (NIMS), Tsukuba 305-0044, Japan}
\date{\today\\[1in]}
\begin{abstract} 
Klein's paradox refers to the transmission of a relativistic particle through a high potential barrier. 
Although it has a simple resolution in terms of particle-to-antiparticle tunneling (Klein tunneling), debates on its physical meaning seem lasting partially due to the lack of direct experimental verification.
In this article, we point out that honeycomb-type photonic crystals (PhCs) provide an ideal platform to investigate the nature of Klein tunneling, where the effective Dirac mass can be tuned in a relatively easy way from a positive value (trivial PhC) to a negative value (topological PhC) via a zero-mass case (PhC graphene). 
Especially, we show that analysis of the transmission between domains with opposite Dirac masses---a case hardly be treated within the scheme available so far---sheds new light on the understanding of the Klein tunneling.
\end{abstract}
\maketitle
\newpage
\section{Introduction}
\noindent

In relativistic quantum mechanics, a potential barrier can become nearly transparent to an incoming particle if the potential exceeds the particle's mass, in stark contrast to the non-relativistic quantum mechanics where a particle cannot transmit such a high potential barrier. This counterintuitive result has been known as Klein's paradox \cite{Klein, Calogeracos}.

\begin{figure}
  \centering\includegraphics[width=8cm]{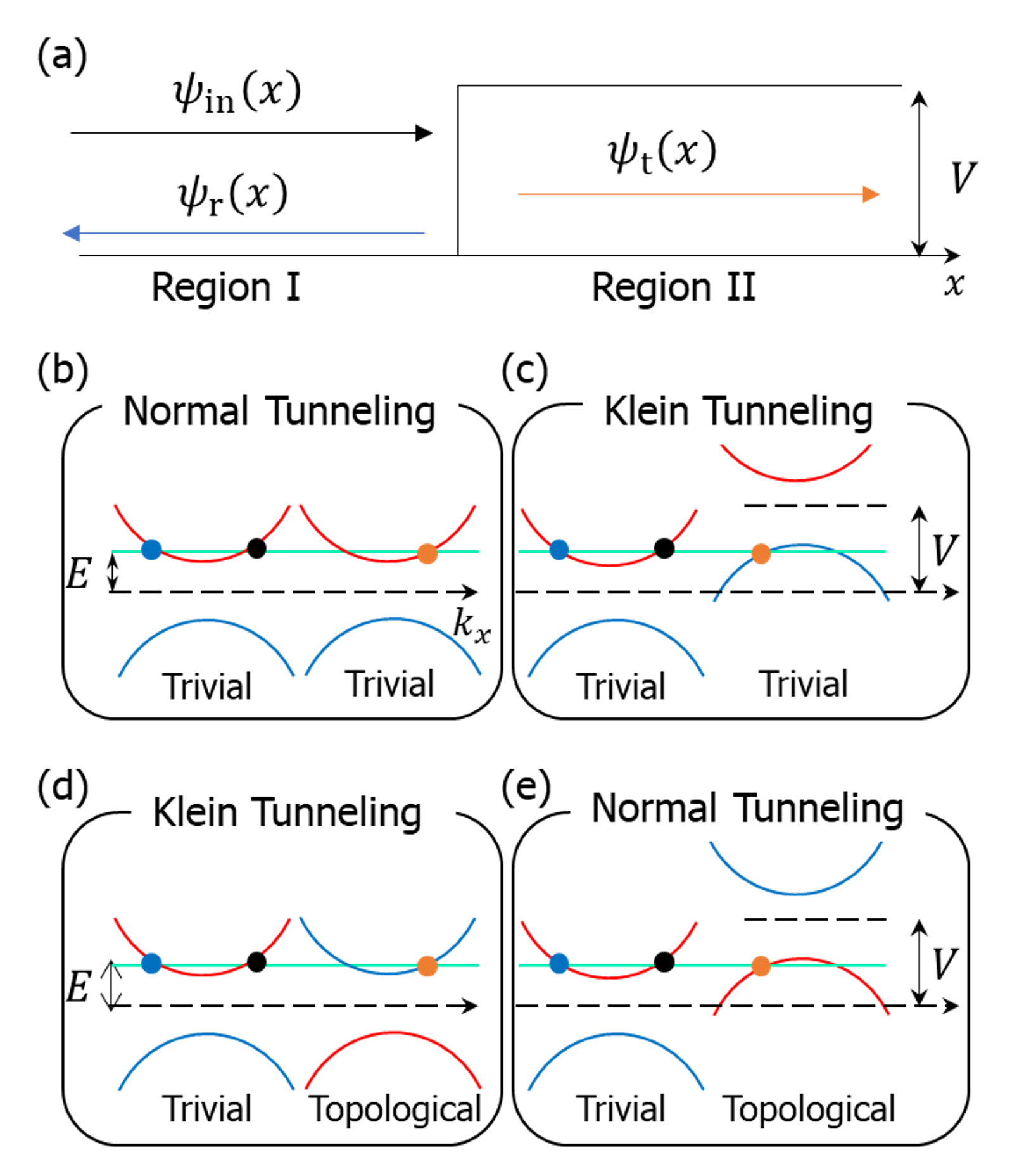}
  \caption{(a) Schematic of the transmission process: A relativistic particle in Region I with mass $m$ and energy $E$ is transmitted into a potential barrier $V$ (Region II). $\psi^\mathrm{in}$, $\psi^\mathrm{r}$, and $\psi^\mathrm{t}$ denote the incident, reflected, and transmitted wavefunctions, respectively. The way of transmissions depends on the value of the potential, namely $V<2mc^2$ (small) or $V>2mc^2$ (large), and the type of PhC in Region II (trivial or topological). (b) Conventional (normal) tunneling at small $V$. The red curve is the ``particle'' band and the blue curve is the ``antiparticle'' band. (c) Massive Klein tunneling at large $V$. (d), (e) Band inversion occurs in topological PhC with negative mass, which allows Klein tunneling at small $V$ at the trivial-topological interface.}
  \label{Fig_Resolution}
\end{figure}

To be explicit, we consider the case that a relativistic particle with mass $m$ and energy $E>0$ is transmitted from a region without potential (Region I) into a potential barrier $V\geq0$ (Region II), as shown in Fig. \ref{Fig_Resolution} (a). The transmission is categorized into three regimes, which we call the small-$V$ regime ($E\geq V$), the reflected regime ($E<V<E+mc^2$), and the large-$V$ regime ($V\geq E+mc^2$). In the small-$V$ regime, the particle is transmitted similarly to the non-relativistic quantum tunneling (Fig. \ref{Fig_Resolution} (b)). In the reflected regime the particle is fully reflected. Most interestingly, in the large-$V$ regime, the particle is transmitted as an antiparticle (Figure \ref{Fig_Resolution} (c)).

Although Klein's paradox has a simple resolution as shown in Fig. \ref{Fig_Resolution} (c), the physical interpretation of Klein tunneling seems still under debate \cite{Hansen, MANOGUE1988, Holstein1998, Calogeracos, Nitta1999, Krekora2004, Krekora2005, Guney2009, Solnyshkov2016, Wang2020}. The large potential energy ($V>2mc^2\approx1$MeV for electron) has led to theoretical interpretations of the paradox in terms of electron-positron pair creation using quantum field theory or quantum electrodynamics, which meanwhile makes its direct verification challenging in experiments of elementary particle physics.
So far, Klein tunneling has been reported experimentally with \emph{massless} particles in various condensed-matter systems \cite{Geim, Beenakker2008, Allain_2011, Huard2007, Gorbachev2008, Stander2009, Young2009, Ni, Jiang}, where there is no strict distinction between particles and antiparticles. In addition, these experiments consider the dispersion near the $\mathrm K$ and $\mathrm K'$ points with finite momenta instead of the $\Gamma$ point, which cannot be considered as ideal platform to clarify the Klein physics in a complete way. Klein tunneling was also studied in (but not limited to) deformed hexagonal lattices \cite{Bahat-Treidel, Garcia}, photonic crystals \cite{Jiang, Longhi, Ni, Ozawa}, and magnonic systems \cite{Duine}, but to the best of our knowledge a direct observation of massive Klein tunneling is still missing.

In this article, we propose that honeycomb-type photonic crystals (PhCs) are ideal systems for investigating massive Klein tunneling. These systems possess doubly degenerate relativistic dispersions near the $\Gamma$ point \cite{Wu-Hu2015, Wu-Hu2016, Kariyado2017, Xie2018, Ozawa2019, Wang2020Review}. So, electromagnetic modes in these systems behave as massive Dirac quasiparticles with four-component spinor wavefunctions.
Recipes of PhC design giving quasiparticles with positive mass (trivial PhC), massless (photonic graphene), and even negative mass (topological PhC) have been established \cite{Wu-Hu2015, Wu-Hu2016, Kariyado2017, Xie2018, Ozawa2019, Wang2020Review}. The advantage of these PhC systems is that the photonic band gap (mass gap) is on the order of $0.1$eV, which can be realized by semiconductor nanofabrication.
We propose that an analog of massive Klein tunneling without potential can appear at the interface of PhCs with positive and negative mass (Fig. \ref{Fig_Resolution} (d), (e)). We show that interfaces between PhC with opposite masses allow us to investigate the essential difference between normal and Klein tunneling.

This article is organized as follows. In Section \ref{Section_Review}, we explain how photonic eigenstates can be described as massive Dirac quasiparticles. In Section \ref{Section_Interface}, we present our model of PhC interface and confirm massive Klein tunneling at the trivial-trivial PhC interface. We reveal that when the particle has a normal incidence, the transmission coefficient through a trivial-trivial PhC interface with a large/small $V$ is identical to that of a trivial-topological interface with a small/large $V$.
In Section \ref{Section_refraction}, we consider the angle dependence of the transmission and find that transmission with a negative index of refraction is achieved in the large-$V$ regime, both for the trivial-trivial and trivial-topological interfaces.
In Section \ref{Section_Soliton}, we investigate whether topological interfacial states \cite{Jackiw-Rebbi, Ozawa2019, Sun2021} disturb the transmission process. In Section \ref{Section_Discussion}, we discuss the implications of our results.

\section{Trivial and Topological Photonic Crystals}
\label{Section_Review}
\begin{figure}
  \centering\includegraphics[width=\linewidth]{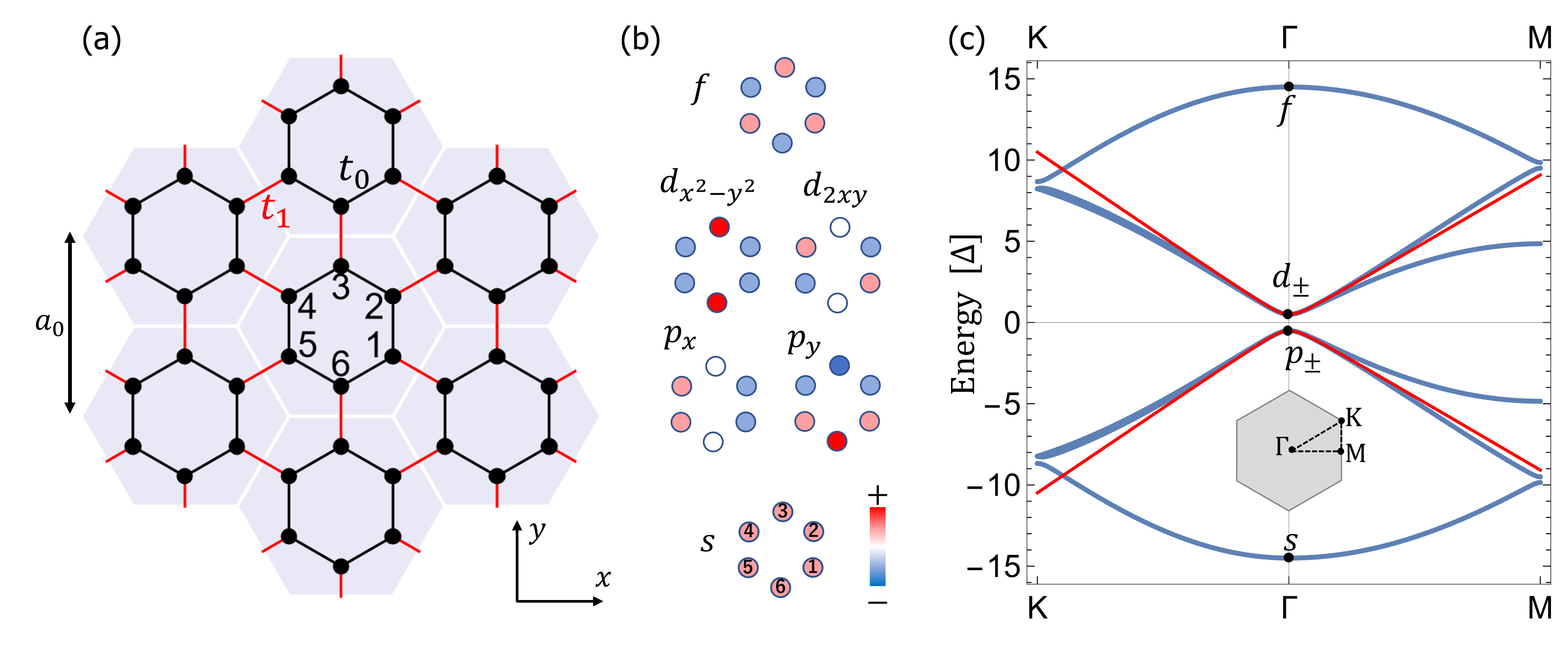}
  \caption{(a) PhC with a honeycomb lattice and hexagonal unit cells which contain six sites. The trivial and topological PhCs are realized by changing the hopping integrals inside and between the unit cells. (b) Photonic eigenstates at the $\Gamma$ point. (c) Photonic dispersion for a trivial PhC with $t_0=1.1t_1$, $M=t_0-t_1$ and $\Delta=2M$. The blue curve is obtained from the tight-binding Hamiltonian Eq. \eqref{TBH}, while the red curve is obtained from the Dirac Hamiltonian Eq. \eqref{DiracH}.}
  \label{Fig_Structure}
\end{figure}
Let us consider a PhC with a honeycomb lattice and choose hexagonal unit cells which contain six sites as shown in Fig. \ref{Fig_Structure} (a). This system can be described by the tight-binding Hamiltonian
\begin{align}
  H_\mathrm{TB}=-t_0\sum_{<i,j>}\ket{i}\bra{j}-t_1\sum_{<i',j'>}\ket{i'}\bra{j'},\label{TBH}
\end{align}
where $t_0,t_1>0$ represent the nearest-neighbor hopping integral inside and between unit cells, respectively. $\ket{i}$ $(i=1\cdots 6)$ represent the position basis inside a unit cell, for which the photonic eigenstates and eigenvalues satisfying the eigenvalue equation
\begin{align}
  H_\mathrm{TB}\ket{\psi}=E\ket{\psi}\label{TBEq}
\end{align}
are well known \cite{Wu-Hu2015, Wu-Hu2016}. The photonic eigenstates at the $\Gamma$ point correspond to the two-dimensional irreducible representation of the $C_{6v}$ point group, namely $\ket{s}$, $\ket{p_x}$, $\ket{p_y}$, $\ket{d_{x^2-y^2}}$, $\ket{d_{2xy}}$ and $\ket{f}$ which are depicted in Fig. \ref{Fig_Structure} (b). Here, the eigenvalue $E=\omega^2/c^2$, with $\omega$ the angular frequency of light and $c$  the speed of light in vacuum, is referred to as the ``energy'' associated with the tight-binding Hamiltonian \eqref{TBH} \cite{Wu-Hu2015}. The states $(\ket{p_x},\ket{p_y})$ and $(\ket{d_{x^2-y^2}},\ket{d_{2xy}})$ are degenerate, so we can define the chiral states $\ket{p_\pm}=\ket{p_x}\pm i\ket{p_y}$ and $\ket{d_\pm}=\ket{d_{x^2-y^2}}\pm i\ket{d_{2xy}}$ which form a new set of basis states.

The photonic eigenstates near the $\Gamma$ point can be obtained from the $\mathbf{k}\cdot\mathbf{p}$ expansion with the basis $[p_+,d_+,p_-,d_-]$ \cite{Wu-Hu2015, Wu-Hu2016} 
\begin{align}
  \hat H=
  \begin{bmatrix}
    \hat H_+&0
    \\
    0&\hat H_-
  \end{bmatrix}
  ,\qquad
  \hat H_\pm=
  \begin{bmatrix}
    -M&\mp iA{k}_\pm\\
    \pm iA{k}_\mp&M
  \end{bmatrix},\label{DiracH}
\end{align}
where ${\mathbf k}=(k_x,k_y)$ describe the wavevector near the band gap, $k_\pm=k_x\pm ik_y$, $M=t_0-t_1$ and $A=a_0t_1/2$. Here, we kept the terms linear in $k_x,k_y$ and ignored higher order terms. As can be seen from Eq. \eqref{DiracH}, the pseudospin-up sector and the pseudospin-down sector are decoupled. Therefore, hereafter we consider only the pseudospin-up sector for simplicity. The pseudospin-up Hamiltonian can be described in terms of Pauli matrices
\begin{align}
  \hat H_+&=M\beta+A(k_x\alpha_1+k_y\alpha_2),
\end{align}
with
\begin{align}
  \renewcommand\arraystretch{0.8}
  \beta   &=\begin{bmatrix}-1&0\\0&1\end{bmatrix},
  \alpha_1=\begin{bmatrix}0&-i\\i&\phantom{-}0\end{bmatrix},
  \alpha_2=\begin{bmatrix}0&\phantom{-}1\\1&\phantom{-}0\end{bmatrix}.\label{PauliMat}
\end{align}
These matrices satisfy the following anti-commutation relations
\begin{align}
  \beta^2=\alpha_1^2=\alpha_2^2=\mymathbb{1},\label{rel1}
  \\
  \beta\alpha_1+\alpha_1\beta=\beta\alpha_2+\alpha_2\beta=\alpha_1\alpha_2+\alpha_2\alpha_1=0,\label{rel0}
\end{align}
where $\mymathbb{1}$ is the $2\times 2$ unit matrix. The photonic eigenstates in the pseudospin-up sector are two-component spinor wavefunctions of the form $\psi_+(\mathbf{r})=[\psi_{p_+}(\mathbf{r}),\psi_{d_+}(\mathbf{r})]^\mathrm{T}$,  with $\mathbf{r}=(x,y)$, which satisfy the eigenvalue equation
\begin{align}
  \hat H_+\ket{\psi_+}=E\ket{\psi_+}.\label{pmEq}
\end{align}
The photonic dispersion is shown in Fig. \ref{Fig_Structure} (c), where the mass gap is equal to $2M$. The blue curves are obtained by solving Eq. \eqref{TBEq} numerically and the red curves show the Dirac dispersion $E=\pm\sqrt{A^2(k_x^2+k_y^2)+M^2}$ which is obtained by solving Eq. \eqref{pmEq} with the plane-wave solution $\psi_+(\mathbf{r})=\psi_+ e^{i\mathbf{k}\cdot\mathbf{r}}$. The blue and red curves coincide near the $\Gamma$ point, which implies that photonic eigenstates can be described as massive Dirac quasiparticles. 

From its definition, it is clear that $M$ can be either positive or negative depending on the values of $t_0$ and $t_1$. A trivial PhC is obtained when $t_0>t_1$ (i.e. $M>0$). On the other hand, a topological PhC is obtained when $t_0<t_1$ (i.e. $M<0$). The negative mass for the topological PhC leads to band inversion, namely exchanging the $\ket{d_+}$ and $\ket{p_+}$ eigenstates in the order of energy near the $\Gamma$ point \cite{Wu-Hu2015, Wu-Hu2016}. In other words, the positive and negative energy states, which correspond to ``particles'' and ``antiparticles'', are inverted in a topological PhC.

In the above, we have shown how photonic eigenstates in honeycomb-type PhC can be described as massive Dirac quasiparticles with positive and negative masses. In what follows,  we use such photonic eigenstates to study massive Klein tunneling at PhC interfaces.

\section{Transmission at {PhC} Interface}
\label{Section_Interface}
\subsection{Model}
We consider the transmission of light through the interface of two PhCs with a potential difference (Figure \ref{Fig_Resolution}), which can be achieved by changing the effective permittivity of the PhC in Region II. The two PhCs have different $M$ and $A$ values, and the system close to the PhC interface is described by the following Hamiltonian
\begin{equation}
  \hat H_+(x)=M(x)\beta-iA(x)\left(\partial_x\alpha_1+\partial_y\alpha_2\right)+V(x)\mymathbb{1},
  \label{Eqn_model}
\end{equation}
where $\hat{\mathbf k}=-i\nabla=-i(\partial_x,\partial_y)$ has been considered, and
\begin{equation}
  M(x)=\begin{cases}
    M^<,&x<0\\M^>,&x>0
  \end{cases},\quad
  A(x)=\begin{cases}
    A^<,&x<0\\A^>,&x>0
  \end{cases},\quad
  V(x)=\begin{cases}
    0,&x<0\\V,&x>0
  \end{cases}.
\end{equation}
We use this Hamiltonian to solve the following eigenvalue equation
\begin{equation}
  \hat H_+(x)\psi_+(\mathbf r)=E\psi_+(\mathbf r),\label{EE}
\end{equation}
where $\psi_+(\mathbf{r})=[\psi_{p_+}(\mathbf{r}),\psi_{d_+}(\mathbf{r})]^\mathrm{T}$. Equation \eqref{EE} gives two coupled differential equations. On the other hand, note that $\hat H_+(x)-V(x)\mymathbb{1}$ is a traceless Hermitian operator. In general, any $2\times2$ traceless Hermitian operator $\hat O$ can be written using Pauli matrices \eqref{PauliMat}. Hermicity implies that the eigenvalues of $\hat O$ are real, so the square of these eigenvalues are always positive. In other words, Hermicity guarantees that the eigenvalues come in positive and negative pairs, and the square of a $2\times2$ traceless Hermitian operator is a diagonal matrix with the same elements. This can be checked using Eq. \eqref{rel1} and Eq. \eqref{rel0}. Explicitly, we obtain
\begin{align*}
  \left[\hat H_+(x)-V(x)\mymathbb{1}\right]^2&=\left[M(x)^2-A(x)^2\nabla^2\right]\mymathbb{1}.
\end{align*}
Using $\mymathbb{1}\psi_+(\mathbf{r})=\psi_+(\mathbf{r})$ and Eq. \eqref{EE}, we arrive at the following decoupled differential equation
\begin{align}
  \left[M(x)^2-A(x)^2\nabla^2\right]\psi_{p_+,d_+}(\mathbf r)=[E-V(x)]^2\psi_{p_+,d_+}(\mathbf{r}).
  \label{Eqn_squared}
\end{align}
We use Eq. \eqref{Eqn_squared} to calculate the eigenvalues while Eq. \eqref{EE} to calculate the eigenstates.
For  the transmission problem in Fig. \ref{Fig_Resolution}, the following plane-wave solution is considered
\begin{align}
  \psi_+(\mathbf{r})&=
  \begin{cases}
    \psi_+^<(\mathbf{r}),&x<0\\\psi_+^>(\mathbf{r}),&x>0
  \end{cases},
  \\
  \psi_+^<(\mathbf r)&=\left[\begin{matrix}\psi_{p_+}^\mathrm{in}\\\psi_{d_+}^\mathrm{in}\end{matrix}\right]e^{i\mathbf{k}^\mathrm{in}\cdot\mathbf{r}}+\left[\begin{matrix}\psi_{p_+}^\mathrm{r}\\\psi_{d_+}^\mathrm{t}\end{matrix}\right]e^{i\mathbf{k}^\mathrm{r}\cdot\mathbf{r}},\quad \psi_+^>(\mathbf r)=\left[\begin{matrix}\psi_+^\mathrm{t}\\\psi_{d_+}^\mathrm{t}\end{matrix}\right]e^{i\mathbf{k}^\mathrm{t}\cdot\mathbf{r}}.
  \label{ansatz1}
\end{align}
Here, $\mathbf{k}^\mathrm{in}=(k_x^\mathrm{in},k_y^\mathrm{in}), \mathbf{k}^\mathrm{r}=(k_x^\mathrm{r},k_y^\mathrm{r})$ and $\mathbf{k}^\mathrm{t}=(k_x^\mathrm{t},k_y^\mathrm{t})$ are the wavevectors of the incident, reflected and transmitted wavefunctions, respectively. By definition of reflection we have $\mathbf{k}^\mathrm{in}\ne\mathbf{k}^\mathrm{r}$ (see Section \ref{Section_refraction} for more detail.)
We solve Eq. \eqref{Eqn_squared} in each of the two regions separately and obtain
\begin{align}
  E=\pm\sqrt{{A^<}^2\lvert{\mathbf{k}^\mathrm{in}}\rvert^2+{M^<}^2}=\pm\sqrt{{A^<}^2\lvert{\mathbf{k}^\mathrm{r}}\rvert^2+{M^<}^2}\label{Eqn_Dispersionin}
\end{align}
for the incident and reflected wavefunctions and
\begin{align}
  E=V\pm\sqrt{{A^>}^2\lvert{\mathbf{k}^\mathrm{t}}\rvert^2+{M^>}^2}\label{Eqn_Dispersiont}
\end{align}
for the transmitted wavefunction.

The continuity of the wavefunction at the interface ($x=0$) implies
\begin{align}
  \psi_{p_+}^\mathrm{in}+\psi_{p_+}^\mathrm{r}=\psi_{p_+}^\mathrm{t},\quad
  \psi_{d_+}^\mathrm{in}+\psi_{d_+}^\mathrm{r}=\psi_{d_+}^\mathrm{t},\quad
  k_y^\mathrm{in}=k_y^\mathrm{r}=k_y^\mathrm{t}.\label{continuity}
\end{align}
Moreover, from Eq. \eqref{EE} with Eq. \eqref{ansatz1}, we obtain
\begin{align}
  \psi_{d_+}^\mathrm{in}=\mykappa^<e^{-i\phi^\mathrm{in}}\psi_{p_+}^\mathrm{in},\quad
  \psi_{d_+}^\mathrm{r}=\mykappa^<e^{-i\phi^\mathrm{r}}\psi_{p_+}^\mathrm{r},\quad
  \psi_{d_+}^\mathrm{t}=\mykappa^>e^{-i\phi^\mathrm{t}}\psi_{p_+}^\mathrm{t},
  \label{vsol_exp}
\end{align}
where
\begin{align}
  \mykappa^<=i\sqrt{\frac{E+M^<}{E-M^<}},\quad \mykappa^>=i\sqrt{\frac{\left(E-V\right)+M^>}{\left(E-V\right)-M^>}}
\end{align}
and $\phi^\mu=\tan^{-1}k_y^\mu/k_x^\mu$. From the conditions $k_y^\mathrm{in}=k_y^\mathrm{r}$ (Eq. (17)) and $\lvert \mathbf{k}^\mathrm{in}\rvert^2=\lvert \mathbf{k}^\mathrm{r}\rvert^2$ (which follows from Eq. \eqref{Eqn_Dispersionin}), we obtain $k_x^\mathrm{r}=-k_x^\mathrm{in}$, i.e. $\phi^\mathrm{r}=\pi-\phi^\mathrm{in}$. Then, from Eq. \eqref{continuity} and Eq. \eqref{vsol_exp}, we obtain the following solution 
\begin{equation}
  \psi_{p_+}^\mathrm{r}=r\psi_{p_+}^\mathrm{in},\quad 
  \psi_{p_+}^\mathrm{t}=t\psi_{p_+}^\mathrm{in},
  \label{Eqn_usol}
\end{equation}
where
\begin{align}
  r&=\frac{\mykappa e^{-i\phi^\mathrm{in}}-e^{-i\phi^\mathrm{t}}}{\mykappa e^{i\phi^\mathrm{in}}+e^{-i\phi^\mathrm{t}}},\quad
  t=\frac{2\mykappa\cos\phi^\mathrm{in}}{\mykappa e^{i\phi^\mathrm{in}}+e^{-i\phi^\mathrm{t}}},
\end{align} 
and $\mykappa=\mykappa^</\mykappa^>$ is the kinematic factor, which agrees with that in Ref. \cite{Calogeracos}.
\subsection{The Transmission Coefficient}
To discuss the transmission and reflection properties quantitatively, we calculate the conserved current $\mathbf{j}^\mu=(j_x^\mu,j_y^\mu)$ which is obtained from the continuity equation
\begin{align}
  \partial_t\left[\psi^{\mu\dagger}\psi^\mu\right]=-\nabla\cdot\mathbf{j}^\mu,
\end{align}
where $k^2=\mathbf{k}\cdot\mathbf{k}$ with $\mathbf{k}=\mathbf{k}^\mu,\mu\in\{\mathrm{in,r,t}\}$.  Note that we do not sum over indices.
Using the time-dependent Dirac equation $i\hbar\partial_t\psi_+=\hat H(x)\psi_+$ with Eq. \eqref{Eqn_model}, \eqref{ansatz1} and \eqref{Eqn_usol} we obtain
\begin{align}
  (j_x^\mathrm{\: in},j_y^\mathrm{\: in})&=\frac{2A^<}{\hbar}\left\lvert\psi_{p_+}^\mathrm{in}\right\rvert\left\lvert\psi_{d_+}^\mathrm{in}\right\rvert\left(\cos\phi^\mathrm{in},\sin\phi^\mathrm{in}\right),
  \\
  (j_x^\mathrm{\: r},j_y^\mathrm{\: r})&=|r|^2(-j_x^{\: \mathrm{in}},j_y^{\: \mathrm{in}}),
  \\
  (j_x^\mathrm{\: t},j_y^\mathrm{\: t})&=\frac{|t|^2}{\mykappa}\left(\frac{\cos\phi^\mathrm{t}}{\cos\phi^\mathrm{in}}j_x^{\: \mathrm{in}},\frac{\sin\phi^\mathrm{t}}{\sin\phi^\mathrm{in}}j_y^{\: \mathrm{in}}\right).
\end{align}
The reflection coefficient $R$ and the transmission coefficient $T$ are calculated from the conserved currents:
\begin{align}
  R&=-\frac{j_x^\mathrm{\: r}}{j_x^\mathrm{\: in}}=\left\lvert  r\right\rvert^2
  =\left\lvert\frac{\mykappa e^{-i\phi^\mathrm{in}}-e^{-i\phi^\mathrm{t}}}{\mykappa e^{i\phi^\mathrm{in}}+e^{-i\phi^\mathrm{t}}}\right\rvert^2,\label{R_Result}
  \\
  T&=\frac{j_x^\mathrm{\: t}}{j_x^\mathrm{\: in}}=\frac{|t|^2}{\mykappa}\frac{\cos\phi^\mathrm{t}}{\cos\phi^\mathrm{in}}=\frac{4\mykappa\cos\phi^\mathrm{t}\cos\phi^\mathrm{in}}{\lvert\mykappa e^{i\phi^\mathrm{in}}+e^{-i\phi^\mathrm{t}}\rvert^2}.\label{T_Result}
\end{align}
It is clear that the condition $R+T=1$ is fulfilled. Although expressions similar to Eq. \eqref{R_Result} and Eq. \eqref{T_Result} have been obtained in literature \cite{Guney2009, Allain_2011}, our results apply for positive and negative masses.
\begin{figure}[t]
  \centering\includegraphics[width=\linewidth]{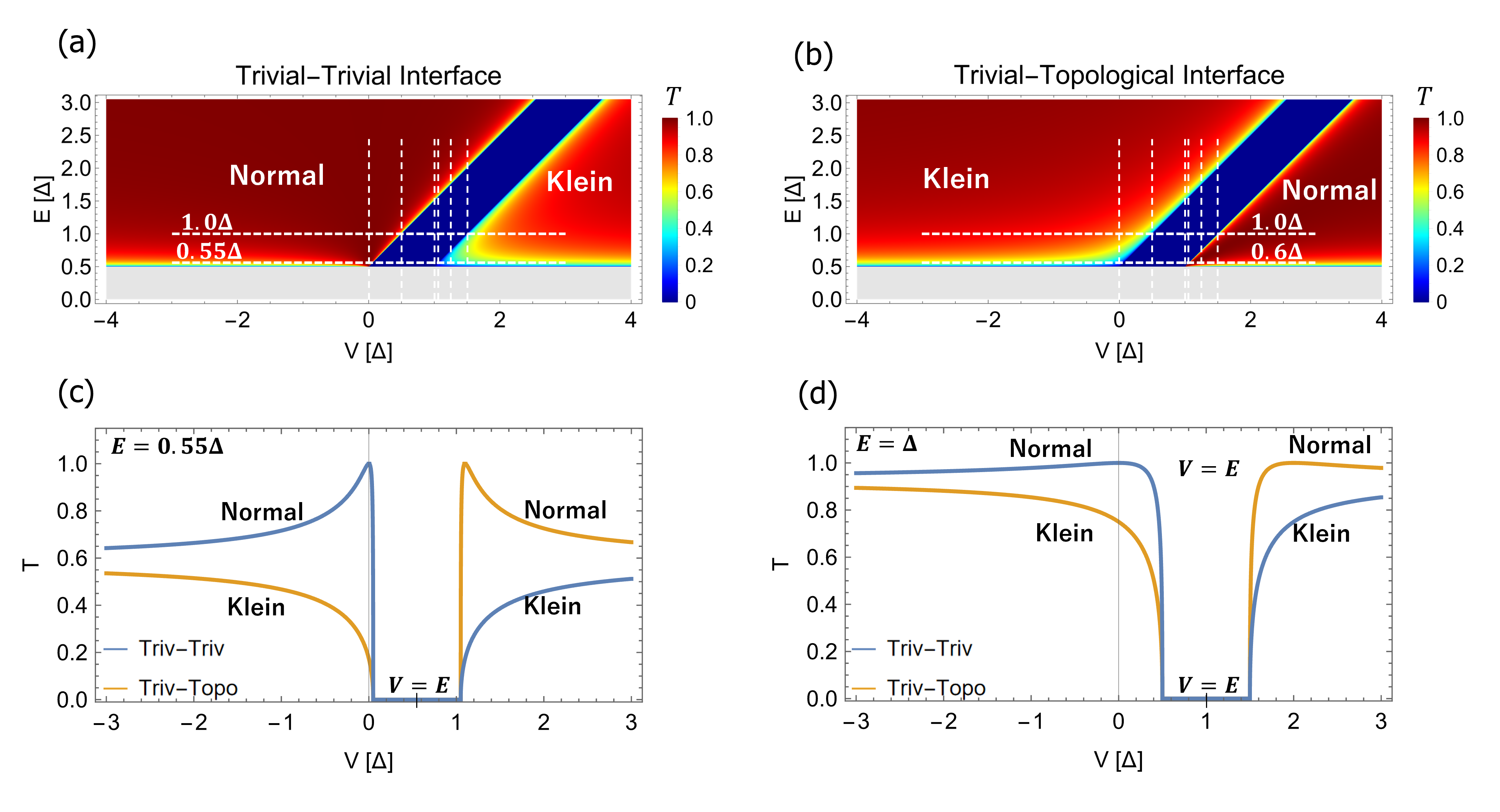}
  \caption{Transmission coefficient $T$ at normal insidence $\phi^\mathrm{in}=0$ for the trivial-trivial interface (a) and for the trivial-topological interface (b).  Here, the bandgap in Region I ($-0.5\Delta\leq E\leq0.5\Delta$ shown by the horizontal gray stripe) and the bandgap in Region II (blue stripe with $T=0$) are both given by $\Delta=2M$ with $M=|M^<|=|M^>|=0.1t_0$. (c) and (d) Line profile of $T$ at $E=0.55\Delta$ and $E=\Delta$. Klein tunneling and normal tunneling are assigned as in Fig. \ref{Fig_Resolution}. Line profile of $T$ at constant $V$ values (vertical dashed lines in (a) and (b)) is shown in Fig. \ref{Fig_Klein_E}.}
  \label{Fig_Klein}
\end{figure}

Figure \ref{Fig_Klein} (a) and (b) show the transmission coefficient $T$ for a trivial-trivial interface and a trivial-topological interface, respectively, with $|M^<|=|M^>|=0.1t_0$ and $\phi^\mathrm{in}=0$. The horizontal gray stripe shows the bandgap $\Delta=2\lvert M^<\rvert$ of the PhC in Region I, i.e. the region without incident particles. The blue stripe is the region with total reflection where $E$ lies within the bandgap of the PhC in Region II. Figure \ref{Fig_Klein} (c) and (d) show the line profile of $T$ at different $E$ values, namely $E=0.55\Delta$ and $E=\Delta$. The blue curve is the tunneling through the trivial-trivial interface, where the large-$V$ regime ($V>E+\Delta/2$) is identical to the Klein tunneling known so far \cite{Klein, Calogeracos}. The orange curve is the tunneling through the trivial-topological interface, which is strikingly different from the blue curve.

To understand the difference between tunnelings at the trivial-trivial and trivial-topological interfaces, we observe that the kinematic factor $\mykappa$ can be written in the following form
\begin{equation}
  \mykappa=\sqrt{\frac{E+M^<}{E-M^<}}\sqrt{\frac{\lvert E-V\rvert-\mathrm{sgn}(E-V)M^>}{\lvert E-V\rvert+\mathrm{sgn}(E-V)M^>}}.\label{etarev}
\end{equation}
We emphasize once again that, if $M^<$ and $M^>$ are both positive, Eq. \eqref{etarev} agrees with that in Ref. \cite{Calogeracos}. However, our result applies for positive and negative masses. The trivial-trivial and the trivial-topological interfaces can be compared by introducing the effective mass $M_{\mathrm{eff}}=\mathrm{sgn}(E-V)M^>$. For the trivial-trivial interface, we have $M_{\mathrm{eff}}>0$ in the small-$V$ regime and $M_{\mathrm{eff}}<0$ in the large-$V$ regime. In contrast, for the trivial-topological interface, we have $M_{\mathrm{eff}}<0$ in the small-$V$ regime and $M_{\mathrm{eff}}>0$ in the large-$V$ regime. In other words, the effective mass is positive for the normal tunneling and negative for the Klein tunneling. Therefore, we conclude that the mass sign of the transmitted particle interchanges normal tunneling and Klein tunneling at the trivial-trivial interface and at the trivial-topological interface.

\begin{figure}[t]
  \centering\includegraphics[width=\linewidth]{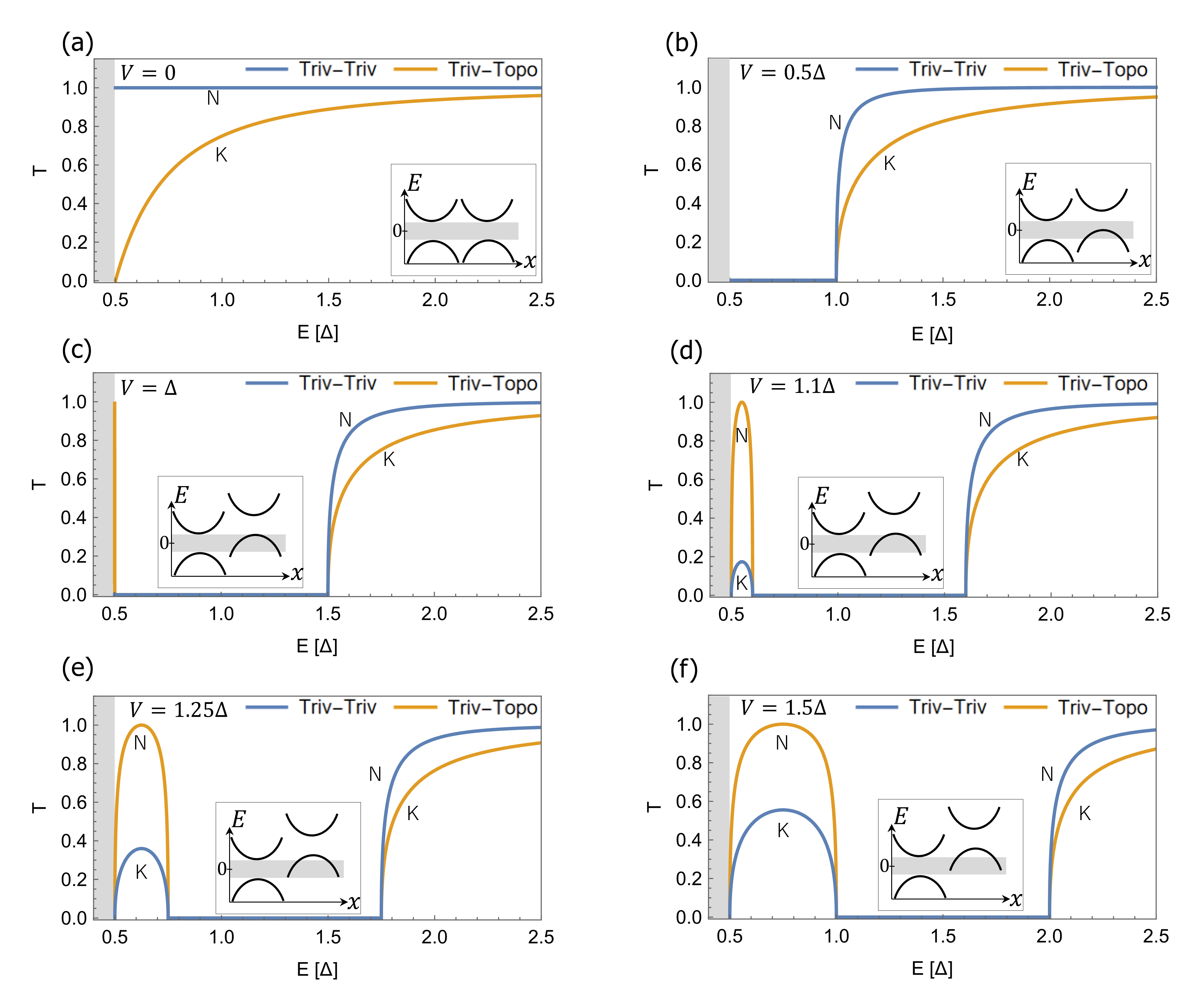}
  \caption{Transmission coefficient $T$ as a function of $E$ at different $V$ values. The gray stripe shows the bandgap of the PhC in Region I. The insets depict the photonic bands at these $V$ values. The blue curve is the tunneling through the trivial-trivial interface with the Klein tunneling appearing at the large-$V$ regime $E\leq V-\Delta/2$. 'N' and 'K' stand for normal tunneling and Klein tunneling, respectively. The orange curve is the tunneling through the trivial-topological interface with the Klein tunneling appearing at the small-$V$ regime $E\geq V+\Delta/2$.}
  \label{Fig_Klein_E}
\end{figure}

In the rest of this section, we investigate the tunneling at fixed $V$ values, which is close to real experimental setups. Figure \ref{Fig_Klein_E} shows the tunneling through the trivial-trivial interface (blue curve) and through the trivial-topological interface (orange curve). Normal tunneling and Klein tunneling are identified as in Fig. \ref{Fig_Klein}. 
At $V=0$ (Fig. \ref{Fig_Klein_E} (a)), the current is fully transmitted at the trivial-trivial interface for any $E\geq\Delta/2$: This result is expected because there is no interface at $V=0$ and the energy spectrum is identical in Region I and Region II. On the other hand, for the trivial-topological interface at $V=0$, the current is fully reflected at the band edge $E=\Delta/2$: This is due to different parities at the $\Gamma$ point induced by band inversion. (This reflection mechanism has been used to construct topological cavity surface emitting lasers \cite{Shao}.) Then, the current is partially transmitted for $E>\Delta/2$ due to the hybridization of $\ket{d}$ and $\ket{p}$ eigenstates.
The transmission changes when $\Delta>$$V>0$ (Fig. \ref{Fig_Klein_E} (b)) because the states in Region I and Region II with the same energy have different hybridization of $\ket{d}$ and $\ket{p}$, i.e. the overlapping between states in each regions is different. As the potential increases above $V=\Delta$ (Fig. \ref{Fig_Klein_E} (c)$\sim$(f)) a dome-like shape appears. The height of the blue dome, which corresponds to the Klein tunneling known so far, increases with $V$. On the other hand, the height of the orange dome does not change with $V$.

\section{Negative index of refraction}
\label{Section_refraction}
Negative index of refraction has been associated with the massive and massless Klein tunneling \cite{Guney2009, Allain_2011}.
Here, we investigate whether this association is still valid for the trivial-topological interface. As in the previous sections we focus on transmission with wavevectors near the $\Gamma$ point.
The physical velocity of a photonic quasiparticle is the group velocity which is defined as $\mathbf{v}_g=\hbar^{-1}\mathrm{grad}_\mathbf{k}E$ \cite{Notomi2000}, and is given from Eq. \eqref{Eqn_Dispersionin} and \eqref{Eqn_Dispersiont}
\begin{equation}
  \mathbf{v}_g(x)=\frac{A(x)^2\mathbf{k}}{\hbar(E-V(x))}.\label{vg}
\end{equation}
Component-wise we have $\mathbf{v}_g=(v_{g,x},v_{g,y})$. Here, we assume that $E$ lies outside of the bandgap such that Eq. \eqref{vg} takes real values. By definition, $v_{g,x}$ is positive for the incident and transmitted states. 
This relation is preserved if $k_x^\mathrm{in}$ is proportional to $\mathrm{sgn}(E)$ and $k_x^\mathrm{t}$ is proportional to $\mathrm{sgn}(E-V)$, which has been depicted in Fig. \ref{Fig_Resolution} (c) and (d). On the other hand, from continuity at the boundary we obtain $k^\mathrm{in}_y=k^\mathrm{t}_y$. If we choose $\phi^\mathrm{in}$ and $E$ to be positive, then $k^\mathrm{in}_y$ and $k^\mathrm{t}_y$ are both positive. If $E-V>0$ then $v_{g,y}$ is positive for the incident and transmitted states, so the index of refraction is positive. On the other hand, if $E-V<0$, then $v_{g,y}$ is positive for the incident state but \emph{negative} for the transmitted state, so the index of refraction is negative. This result can be generalized using Eq. \eqref{vg} and we obtain
\begin{align}
  Q=\frac{v_{g,y}^\mathrm{t}}{v_{g,y}^\mathrm{in}}=\frac{{A^>}^2E}{{A^<}^2(E-V)}\propto\frac{\mathrm{sgn}(E)}{\mathrm{sgn}(E-V)}.
  \label{Eqn_Snell}
\end{align}

\begin{figure}[t]
  \centering\includegraphics[width=0.6\linewidth]{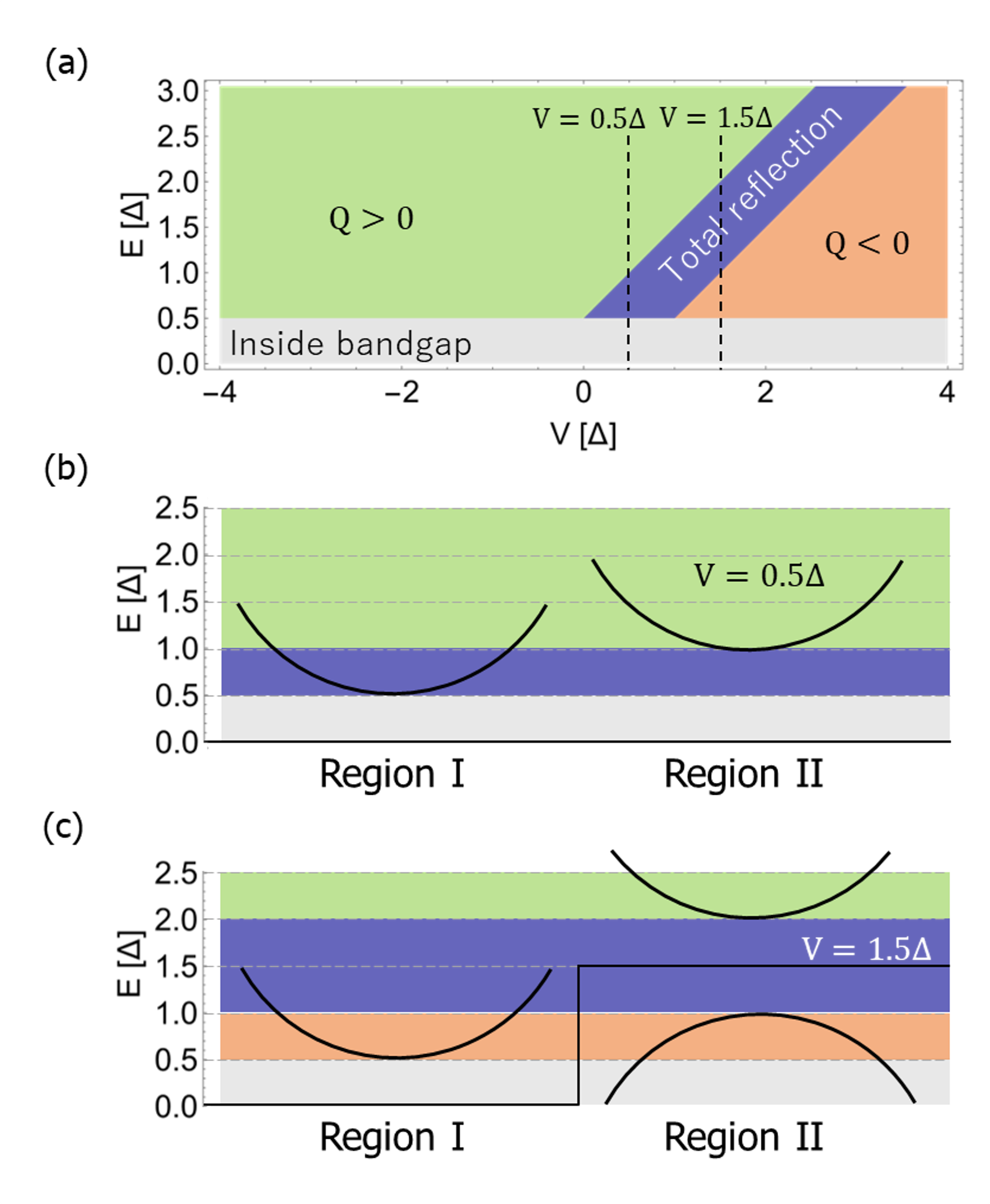}
  \caption{(a) Dependence of the index of refraction on the potential $V$.  The index of refraction is positive in the small-$V$ regime (green region) but negative in the large-$V$ regime (orange region). Note that this result is independent of the type of interface, i.e. independent of the mass sign. (b) and (c) Sign of the index of refraction at $V=0.5\Delta$ and $V=1.5\Delta$, respectively. Negative index of refraction appears for tunneling from a concave-up band to a concave-down band (or vice versa).}
  \label{Fig_Refraction}
\end{figure}

The above equation can be understood as an analog of Snell's law where the index of refraction can be positive or negative depending on the values of $E$ and $V$ (Similar results are obtained in refs. \cite{Guney2009, Allain_2011}), as shown in Fig. \ref{Fig_Refraction}. In particular, we obtain a negative index of refraction in the large-$V$ regime which has $E>0$ and $E-V<0$, i.e. for tunneling from a concave-up band to a concave-down band (Fig. \ref{Fig_Refraction} (c)). This situation is analogous to ref. \cite{Notomi2000} which explains the negative refraction by a concave-down photonic band. Note that Eq. \eqref{Eqn_Snell} is independent of the sign of $M^<$ and $M^>$  since the mass enters into the Hamiltonian as $M^2$. Therefore, negative refraction appears at the large-$V$ regime of both trivial-trivial and trivial-topological interfaces. On the other hand, we have shown in the previous section that for a trivial-topological interface, Klein tunneling appears in the small-$V$ regime while normal tunneling appears in the large-$V$ regime. This result implies that negative refraction is not directly related to massive Klein tunneling.
\section{Jackiw-Rebbi soliton}
\label{Section_Soliton}
It is well known that the Jackiw-Rebbi soliton appears at the center of the bandgap of a positive-negative mass interface \cite{Jackiw-Rebbi}.  Here, we check whether such states reduce the transmission. As in previous studies \cite{Sun2021} we split the Hamiltonian into
\begin{equation}
  \hat H_+=\hat H_{0+}
  +
  \Delta\hat H_+
  \label{Eqn_pert}
\end{equation}
with
\begin{align}
  \hat H_{0+}&=
  \left[
    \begin{matrix}
      -M(x)+V(x)&-A(x)\partial_x\\
      A(x)\partial_x&M(x)+V(x)
    \end{matrix}
  \right],
  \label{Eqn_H0}
  \\
  \Delta\hat H_+&=\left[
    \begin{matrix}
      0&-iA(x)\partial_y\\
      -iA(x)\partial_y&0
    \end{matrix}
    \right]
    \label{Eqn_DeltaH},
\end{align}
where $\Delta\hat H_+$ is taken as a perturbation.
We assume a wavefunction of the form \cite{Sun2021}
\begin{equation}
  \psi_+^<(\mathbf{r})=\left[\begin{matrix}\psi_{p_+}^<\\\psi_{d_+}^<\end{matrix}\right]e^{\kappa^<x+ik_yy},\quad \psi_+^>(\mathbf{r})=\left[\begin{matrix}\psi_{p_+}^>\\\psi_{d_+}^>\end{matrix}\right]e^{-\kappa^>x+ik_yy}.
  \label{JRAnsatz}
\end{equation}
The wavefunction must vanish as $\lvert x\rvert\to\infty$ which requires $\kappa^<$ and $\kappa^>$ to be positive.
After solving $\hat H_+(x)\psi_+(\mathbf r)=E_0\psi_+(\mathbf r)$ with Eq. \eqref{Eqn_H0} and Eq. \eqref{JRAnsatz} we obtain
\begin{align}
  \psi_{d_+}^<=-\frac{A^<\kappa^<}{M^<-E_0}\psi_{p_+}^<,\quad \psi_{d_+}^>=+\frac{A^>\kappa^>}{M^>-(E_0-V)}\psi_{p_+}^>,\label{solitonsol}
  \\
  \kappa^<=\frac{\sqrt{{M^<}^2-E_0^2}}{A^<},\quad \kappa^>=\frac{\sqrt{{M^>}^2-(E_0-V)^2}}{A^>},
\end{align}
where $\kappa^<$ and $\kappa^>$ are real numbers. From the solution of $\kappa^<$ we obtain $-M^<<E_0<M^<$, i.e. $E_0$ must be inside the bandgap of the trivial PhC. On the other hand, from the solution of $\kappa^>$ we obtain $-\lvert M^>\rvert+V<E_0<\lvert M^>\rvert+V$ (recall that $M^><0$), i.e. $E_0$ must also be inside the bandgap of the topological PhC. These conditions are satisfied simultaneously only if $\lvert V\rvert<M^<-M^>$. From the continuity of the wavefunction at $x=0$ we obtain
\begin{align}
  \psi_{p_+}^<&=\psi_{p_+}^>=\psi_{p_+},
  \\
  \psi_{d_+}^<&=\psi_{d_+}^>=\psi_{d_+}.
\end{align}
These conditions are satisfied simultaneously if  $E_0$ takes the following form
\begin{align}
  E_0=\frac{VM^<}{M^<-M^>},\label{JRE}
\end{align}
with $\lvert V\rvert<M^<-M^>$, which generalizes the zero-energy Jackiw-Rebbi soliton to the case with a potential. 
Substituting this expression into Eq. \eqref{solitonsol} we obtain
\begin{align}
  \psi_{d_+}=-\mykappa_s\psi_{p_+}
\end{align}
with
\begin{align}
  \mykappa_s=\sqrt{\frac{(M^<-M^>)+V}{(M^<-M^>)-V}}.
\end{align}
Therefore, the interfacial state is described by the following wavefunction
\begin{equation}
  \psi_+(\mathbf{r})=\psi_{d_+}\begin{bmatrix}-1/\mykappa_s\\1\end{bmatrix}\begin{cases}
    e^{\frac{\sqrt{{M^<}^2-V^2{M^<}^2/(M^<-M^>)^2}}{A^<}x+ik_yy},&x<0
    \\
    e^{-\frac{\sqrt{{M^<}^2-V^2{M^<}^2/(M^<-M^>)^2}}{A^>}x+ik_yy},&x>0
  \end{cases}
\end{equation}
where the normalization condition $\int_{-\infty}^\infty\lvert\psi(x,y)\rvert^2dx=1$ implies $\psi_{d_+}=\sqrt{\frac{\mykappa_s^2}{1+\mykappa_s^2}\frac{2\kappa^<\kappa^>}{\kappa^<+\kappa^>}}$. This solution is stable if $\lvert V \rvert<\lvert M^<-M^>\rvert$, i.e. stable in the small-$V$ and reflected regime with a nonzero common global bandgap. The perturbation Eq. \eqref{Eqn_DeltaH} gives an additional energy
\begin{align}
  \Delta E=\braket{\psi_+|\Delta H_+|\psi_+}=-\frac{2\eta_s}{1+\eta_s^2}\left(\frac{\kappa^<A^>+\kappa^>A^<}{\kappa^<+\kappa^>}\right)k_y
\end{align}
so the energy of the interfacial state is
\begin{align}
  E_\text{interface}=\frac{VM^<}{M^<-M^>}-\frac{2\eta_s}{1+\eta_s^2}\left(\frac{\kappa^<A^>+\kappa^>A^<}{\kappa^<+\kappa^>}\right)k_y\label{E_Interface}.
\end{align}
Note that the stability of the soliton is not affected by this perturbation. 

\begin{figure}[t]
  \centering\includegraphics[width=\linewidth]{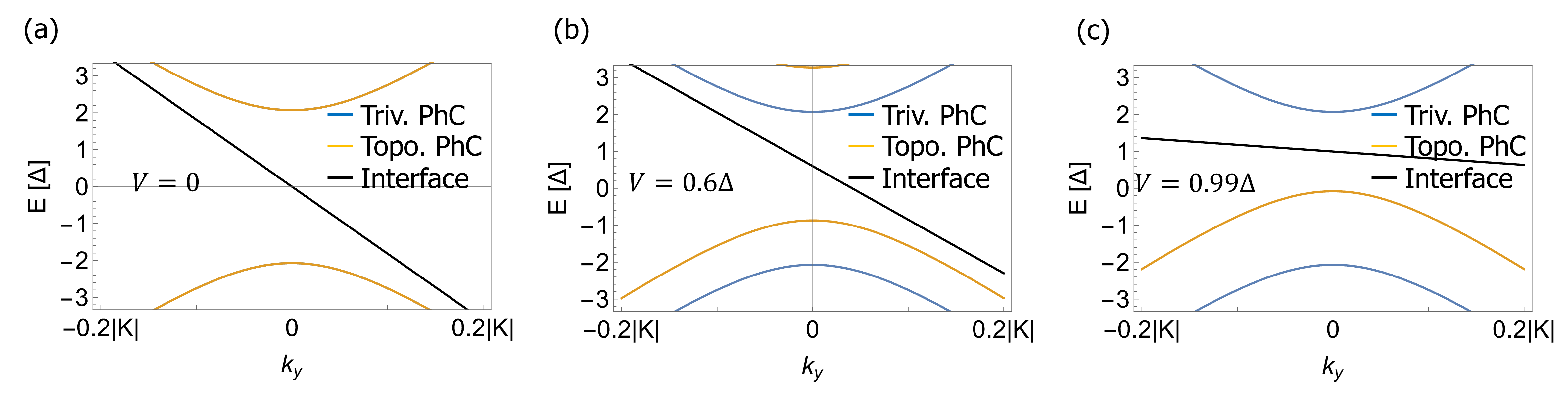}
  \caption{Plot of the bulk band of the trivial PhC (blue curve) and topological PhC (orange curve) as a function of $k_y$ with the interfacial state (black line). Here, we consider only the pseudospin-up states, so only one interfacial state appears due to pseudospin-momentum locking. (a)$\sim$(c): For $k_x>0$ the interfacial state is lower than the transmitted state (upper band of trivial PhC), so the interfacial state does not affect the transmission.}
  \label{Fig_Interface}
\end{figure}

Figure \ref{Fig_Interface} shows the energy of the interfacial state (black line) at different $V$ values. On top of that, we also plot the bulk band of the trivial PhC (blue curve) and topological PhC (orange curve) as a function of $k_y$ with fixed $k_x$ values. Since $\eta_s$, $\kappa^\lessgtr$, $A^\lessgtr$ are positive parameters, the group velocity $v_{g,y}=\hbar^{-1}\partial\Delta E/\partial k_y$ is negative, i.e. the soliton (with up spin) propagates in the negative $y$ direction. For the pseudospin-down sector, the soliton propagates in the positive-$y$ direction, which is a manifestation of pseudospin-momentum locking in topological interfaces with time reversal symmetry \cite{Hasan2010, Qi2011, Wu-Hu2015}. We find that the slope of the interfacial state is reduced as $V$ increases. In the limit $\lvert V\rvert\to\Delta$ the interfacial state becomes flat, which may have interesting future applications.

Since $k_y$ is conserved, transition from the incident state to the Jackiw-Rebbi interfacial state is possible only if both states share the same $E$ and $k_y$ value. Figure \ref{Fig_Interface} (a)$\sim$(c) shows that the dispersion of the interfacial state is always lower than that of the transmitted state (upper band of trivial PhC). Therefore, we conclude that the Jackiw-Rebbi interfacial state does not affect the transmission.

\section{Discussion}
\label{Section_Discussion}
Finally, we summarize our results and discuss their implications.
In this article, we point out that honeycomb-type PhCs provide an ideal platform to investigate the nature of Klein tunneling, where the effective Dirac mass can be tuned in a relatively easy way from a positive value (trivial PhC) to a negative value (topological PhC) via a zero-mass case (PhC graphene). We considered two types of interfaces, namely the trivial-trivial interface and the trivial-topological interface.

First, by studying the transmission at both types of interfaces, we found that transmission of a particle at normal incidence at the trivial-trivial PhC interface with a large/small $V$ is identical to that of a trivial-topological interface with a small/large $V$. The reason for this duality is that in the large-V regime, the mass sign of the transmitted particle is effectively reversed at the trivial-trivial interface. Particle-antiparticle tunneling occurs even without high potential at the trivial-topological interface. Therefore, we conclude that the high potential is not necessary for the definition of Klein tunneling.

Second, we considered the angle dependence of the transmission and found that transmission with a negative index of refraction is achieved in the large-$V$ regime both for the trivial-trivial and trivial-topological interfaces. In fact, it has been shown that negative refraction can be achieved with a photonic band with concave-down curvature \cite{Notomi2000}, which is what we obtain in the large-$V$ regime considered by Klein. While negative index of refraction has been associated with massive and massless Klein tunneling, here we have shown that massive Klein tunneling appears in the small-$V$ regime of a trivial-topological interface. Therefore, the large potential and massive Klein tunneling should be considered separately.

Third, we found that the Jackiw-Rebbi soliton solution at the trivial-topological PhC interface does not disrupt the transmission. Therefore, our results can be tested in PhC interfaces. Our results are not limited to PhC systems but also apply to other Dirac systems.

\vspace{0.5in}
\titlespacing*{\section}
{0pt}{10pt}{0pt}
\section*{Funding}
\noindent This research was funded by the CREST, JST (Core Research for Evolutionary Science and Technology,  Japan Science and Technology Agency) Grant Number JPMJCR18T4.
\section*{Acknowledgments}
\noindent K.N. thank Toyoki Matsuyama and Satoshi Tanda for discussions on Klein tunneling.
\section*{Disclosures}
\noindent The authors declare no conflicts of interest.
\section*{Data Availability}
\noindent The datasets generated during and/or analyzed during the current study are available from the corresponding author on reasonable request.
\newpage
\bibliographystyle{apsrev4-2}
\bibliography{References}
\end{document}